\documentclass[amsmath,amssymb,showpacs,groupedaddress]{revtex4-1}
\usepackage{graphicx}
\usepackage{natbib}%

\begin{document}

\title{Single-File Diffusion of Externally Driven Particles}

\author{Artem Ryabov}
\email{rjabov.a@gmail.com}
\author{Petr Chvosta}
\affiliation{
 Department of Macromolecular Physics,
 Faculty of Mathematics and Physics,
 Charles University in Prague,
 V Hole{\v s}ovi{\v c}k{\' a}ch 2,
 CZ-180~00~Praha, Czech Republic
}

\begin{abstract}
We study 1-D diffusion of $N$ hard-core interacting Brownian particles driven by the space- and time-dependent external force. We give the exact solution of the $N$-particle Smoluchowski diffusion equation. In particular, we investigate the nonequilibrium energetics of two interacting particles under the time-periodic driving. The hard-core interaction induces entropic repulsion which differentiates the energetics of the two particles. We present exact time-asymptotic results which describe the mean energy, the accepted work and heat, and the entropy production of interacting particles and we contrast these quantities against the corresponding ones for the non-interacting particles.
\end{abstract}


\maketitle
\emph{Introduction:} Stochastic dynamics of interacting particles in a one-dimensional environment is both of great practical and theoretical interest. Due to the one-dimensionality of the problem, the inter-particle interactions play a crucial role as they alter qualitative features of the particle dynamics. The type of interaction we deal with in this Letter is a so called \emph{hard-core interaction}.

The diffusion of particles in narrow channels, where the particles cannot pass each other and their relative ordering is conserved, is  known as the \emph{single-file diffusion} (SFD).  The concept of SFD was first introduced by Hodgkin and Keynes in relation to the transport of water and ions through the molecular-sized channels in membranes \cite{Hodgkin55}. Since then, numerous examples of SFD in biological, chemical, and physical processes were studied (e.g. transport of adsorbate molecules through zeolites with a one-dimensional channel system \cite{Hahn96, Karger_zeol}, geometrically constrained nano-sized particles in nano-sized pores \cite{Dekker_nano}, migration of adsorbed molecules on surfaces \cite{HalpinZhang}, diffusion in nanotubes \cite{Carbon, Dipeptide}, a carrier migration in polymers and superionic conductors \cite{vanGool}, diffusion of colloids in one-dimensional channels \cite{Bech2000, Bech2004, inPolym2006}, confined dynamics of millimetric steel balls \cite{Delfau}).

While the global properties of a SFD-system are identical to those for the system of independent particles, the dynamics of an individual particle
(also called \emph{tagged particle} or \emph{tracer}) is considerably different \cite{Rodenbeck98, Kutner81, Kutner83}. Theoretical description of SFD was introduced by Harris in 1965. In his pioneering study \cite{Harris65} he showed that the mean square displacement of the tagged particle increases with time as $t^{1/2}$ (in contrast to its
linear increase for the single free particle). This result was subsequently reestablished by many other authors using different mathematical tools (for a comprehensive review
cf.\ the Introduction in \cite{SFDLizana}). The first exact solutions of the diffusion equation for SFD systems appeared only recently. The solution for an arbitrary number $N$ of identical particles diffusing along the infinite line has been obtained by R{\"o}denbeck et al. in \cite{Rodenbeck98} via the reflection principle. Using a different theoretical procedure, the result has been re-derived in \cite{SFDKumar}.
The exact solution for the diffusion within the finite 1-D interval has been found in \cite{SFDLizana} through the Bethe ansatz. The exact solution for $N=2$ particles
having different diffusion constants has been obtained in \cite{SilbAmbj2008}. Another remarkable exact result for $N=2$ has been derived in \cite{SilbN=2}
where the authors assume the finite-range interaction between the particles. The first calculation including the time-independent external force has been
published in \cite{SilbeyForceField}. Using a so called ``Jepsen line'', the probability density for the single tagged particle was obtained in the limit of an infinite system with the constant particle density. As far as we know, no exact solution of the SFD of $N$ particles driven by the external time-dependent force has been published, yet.

The main objective of this Letter is to analyse the thermodynamic properties of an exactly solvable, experimentally verifiable model of an externally driven SFD. Hence the setting must incorporate the following attributes:
a) the hard-core interaction among the particles,
b) the non-trivial equilibrium state if the external driving is switched off,
c) the time-periodic external driving inducing a periodic response.
We first give the exact solution (\ref{THMN}) of the diffusion equation with an arbitrary time- and space-dependent external driving of the $N$ hard-core interacting particles. The solution emerges in the form of a sum of products of the \emph{single-particle} probability densities which solve the corresponding one-particle diffusion problem. Differently speaking, in the second step, we need an exactly solvable one-particle model which would include the above ingredients b) and c). We focus on the problem which is related with the famous barometric formula. We consider a particle diffusing in a half-space $x\in (0,+\infty)$ and acted upon by the spatially homogeneous and time-dependent force with a reflective boundary placed at $x=0$. Using the general $N$-particle solution and the single-particle probability density for the barometric problem we investigate in detail the diffusion of
two interacting particles.

\emph{General solution:} Consider $N$ identical hard-core interacting Brownian particles
\footnote{In order to incorporate the hard-core interaction in 1-D, the particles can be represented by rods of the length $l$. The hard-core interaction in such system means that the space occupied by one rod is inaccessible to the neighbouring rods. Generally the diffusion of hard rods can be mapped onto the diffusion of \emph{point} particles by the rescaling of space variables (see e.g. \cite{SFDLizana}). Therefore, all our considerations are done for systems of point particles.
},
each with the same diffusion constant $D$, diffusing under the influence of the external force $F(x,t)$.
Let the vector $\vec{x} = x_{1},...\,, x_{N}$ denotes positions of the particles at the time $t$ and let $\vec{y} = y_{1},...\,, y_{N}$ stands for their initial positions at $t_{0}\leq t$. The time evolution of the joint conditional probability density for the positions of the particles, say $p^{(N)}(\vec{x}; t \, | \, \vec{y}; t_{0})$, is controlled
by the Smoluchowski diffusion equation \cite{Risken}
\begin{eqnarray}
 \frac{\partial}{\partial t} p^{(N)}(\vec{x}; t \, | \, \vec{y}; t_{0})   
 = \sum_{j=1}^{N} \left\{ -\frac{\partial}{\partial x_j} \frac{F(x_{j},t)}{m \gamma}  \right. +
\left. D\frac{\partial^2}{\partial x_{j}^{2}}  \right\} p^{(N)}(\vec{x}; t \, | \, \vec{y}; t_{0}) \,\,,
\label{SmolNHC}
\end{eqnarray}
with the initial condition
\begin{equation}
\label{initialN}
p^{(N)}(\vec{x}; t_{0} \, | \, \vec{y}; t_{0}) = \delta (x_1-y_1) \dots \delta (x_N-y_N)\,\,.
\end{equation}
In Eq. (\ref{SmolNHC}) $m$ and $\gamma$ are the mass of and the friction coefficient for the single particle, respectively.
Suppose that the particles are initially ordered as $y_{1}<y_{2}<...<y_{N}$. Owing to their hard-core interaction, the ordering is conserved at all $t\geq t_{0}$ and the
function $p^{(N)}(\vec{x}; t \, | \, \vec{y}; t_{0})$ vanishes outside the domain $x_{1}  <x_{2} < ...\, < x_{N}$.
This restraint is guaranteed by the \emph{non-crossing conditions}
\begin{equation}
\label{non-crossingN}
 \left. \left( \frac{\partial}{\partial x_{j+1}} - \frac{\partial}{\partial x_{j}} \right)
p^{(N)}(\vec{x}; t \, | \, \vec{y}; t_{0}) \right|_{x_{j+1}=x_{j}}
\!\!\!\!\!\!\!\!\!\!
= 0 \,\,,
\end{equation}
$j = 1,...\,,N-1$. In order to derive the conditions, one first introduces the $N$-particle probability current $\vec{J} = (J^{(N)}_{1}, ..., J^{(N)}_{N})$ with the
$j$-th component
\begin{equation}
\label{fluxN}
J^{(N)}_{j}(\vec{x}; t \, | \, \vec{y}; t_{0}) = \\
 \left\{  \frac{1}{m \gamma} F(x_{j},t)- D\frac{\partial}{\partial x_{j}} \right\}
 p^{(N)}(\vec{x}; t \, | \, \vec{y}; t_{0}) \,\,.
\end{equation}
Thereupon the hard-core interaction can be implemented through the requirements
\begin{equation}
\label{flux_cond}
\left. J^{(N)}_{j}(\vec{x}; t \, | \, \vec{y}; t_{0})-J^{(N)}_{j+1}(\vec{x}; t \, | \, \vec{y}; t_{0})\right|_{x_{j+1}=x_{j}}
\!\!\!\!\!\!\!\!\!\!
= 0\,\,.
\end{equation}
Inserting the components (\ref{fluxN}), we arrive at the conditions (\ref{non-crossingN}). Notice that the argumentation holds true only for
the \emph{identical} particles. In other words, the parameters $D$, $m$, $\gamma$, and the external force $F(x,t)$ must be the same for any individual particle
from the SFD system.

Let us now consider the Smoluchowski diffusion equation (\ref{SmolNHC}) with $N=1$. Assume we know the solution $p^{(1)}(x; t \, | \, y; t_{0})$
of this single-particle diffusion problem. Then we claim the following. The exact $N$-particle joint probability density $p^{(N)}(\vec{x}; t \, | \, \vec{y}; t_{0})$ which fulfils (\ref{SmolNHC}), (\ref{initialN}) and (\ref{non-crossingN}) reads
\begin{equation}
 p^{(N)}(\vec{x}; t \, | \, \vec{y}; t_{0}) =
\sum_{k=1}^{N!} p^{(1)}(x_{1}; t \, | \, y_{\pi_{k}(1)}; t_{0}) 
 p^{(1)}(x_{2}; t \, | \, y_{\pi_{k}(2)}; t_{0}) ...\,
p^{(1)}(x_{N}; t \, | \, y_{\pi_{k}(N)}; t_{0})\,\,,
\label{THMN}
\end{equation}
if $x_{1}<x_{2}<...\,<x_{N}$, and it vanishes otherwise. The summation is taken over all permutations $\pi_{k}$. A given permutation specifies the initial conditions of the
single-particle densities on the right hand side. The claim can be proven by a direct check: we simply insert the right hand side of Eq.\ (\ref{THMN}) into (\ref{SmolNHC}) and (\ref{non-crossingN}).

The formula (\ref{THMN}) expresses the \emph{exact} solution of the \emph{many-particle problem with the hard-core interaction} through a simpler object which is
the \emph{single-particle} probability density.
Notice that in the special case of the SFD without the external driving ($F(x,t)=0$) the result (\ref{THMN}) correctly reproduces the solutions
which were previously derived using the reflection principle \cite{Rodenbeck98}, or the Bethe ansatz \cite{SFDLizana}. However, these techniques fail in the presence of the external time-dependent force $F(x,t)$. This is the reason why the SFD under the external driving has remained unsolved, yet.
Our derivation demonstrates that the structure of the solution as given by Eq.\ (\ref{THMN}) remains valid even in the SFD under the external driving.

\emph{Barometric SFD of two particles:}
Place two Brownian particles into the long 1-D channel, plug one end of the channel and let each particle be acted upon by the space-homogeneous and time-dependent force $F(t)$.
In other words, let the two particles diffuse in the external potential
\begin{equation}
\label{2Ppotential}
	\phi(x_{i},t) = \left\{
	\begin{array}{ll}
	\displaystyle -\, x_{i} \, F(t)
			& \,\, \textrm{for} \,\, x_{i}>0\,\,,\\[15pt]
	+\infty & \,\, \textrm{for} \,\, x_{i}<0\,\,, \quad i = 1,2\,\,.
	\end{array}
	\right.
\end{equation}
The driving force incorporates two components,
\begin{equation}
\label{oscildrift}
	F(t) = F_0 + F_{1} \sin(\omega t)\,\,.
\end{equation}
The time-independent component $F_{0}$ alone would push the particles to the left towards the reflective boundary (if $F_{0}<0$), or to the right (if $F_{0}>0$). The time-dependent component $F_{1} \sin(\omega t)$ harmonically oscillates with the angular frequency $\omega$. Having in mind this scenario, we wish to contrast the dynamics and the energetics of the system of two interacting particles against the model without the interaction. In the latter case the analysis trivially follows from the solution concerning the \emph{single-diffusing particle}.

Assuming the driving force (\ref{oscildrift}), the most interesting physics emerges if the oscillating component $F_{1}\sin(\omega t)$ superposes with a negative static force $F_{0} < 0$. In this case, $F_{0}$ acts against the general spreading tendency stemming from the thermal fluctuations of the surroundings. Due to the periodic driving the system of particles approaches a definite \emph{steady state} exhibiting cyclic changes of the probability density regardless the initial positions $y_{1}, y_{2}$. A necessary and sufficient condition for the existence of such steady state is $F_{0}<0$ [24]. Let us denote probability density of the single-diffusing particle in the steady state as $p_{\rm S}(x;t)$ (i.e., in the steady state, $p^{(1)}(x;t\,|\,y,t_{0})\sim p_{\rm S}(x;t)$, and
$p_{\rm S}(x;t+2\pi/\omega)=p_{\rm S}(x;t)$, for any $x$ and $t$). In the steady state,
the right hand side of the formula (\ref{THMN}) collapses into the product of just $N$ terms. Explicitly, for $N=2$, the two-particle joint probability density in the steady state reads
\begin{equation}
\label{P2as}
\widetilde{p}^{(2)}\!(x_{1},x_{2};t) = 2p_{\rm S}(x_{1}\,;t) p_{\rm S}(x_{2}\,;t) \,\,,
\end{equation}
if $x_{1}<x_{2}$, and it vanishes otherwise. Hence the marginal probability densities for the position of the left (L) and the right (R) interacting particle
are
\begin{eqnarray}
\label{PleftAS}
p_{{\rm L}}(x;t) &=& 2 p_{\rm S}(x\,;t)\! \int_{x}^{+\infty}
\!\!\!\!\!{\rm d}x_{2}\, p_{\rm S}(x_{2}\,;t)\,\,,\\
\label{PrightAS}
p_{{\rm R}}(x;t) &=& 2 p_{\rm S}(x\,;t)\! \int_{0}^{x}
{\rm d}x_{1}\, p_{\rm S}(x_{1}\,;t)\,\,,
\end{eqnarray}
respectively. Carrying out the integrations and using the normalization condition for the function $p_{\rm S}(x;t)$, we have proven an important identity
\begin{equation}
\label{sumdens}
p_{\rm R}(x\,;t)+p_{\rm L}(x\,;t)=2 p_{\rm S}(x;t)\,\,.
\end{equation}
For any $t$, there exist a unique coordinate $\xi(t)$ such that $p_{\rm L}(\xi (t);t)= p_{\rm R}(\xi (t);t)$. For $x\lessgtr\xi(t)$, $p_{\rm R}(x;t)\lessgtr p_{\rm L}(x;t)$.

Being a periodic function of time, the probability density $p_{\rm S}(x;t)$ can be represented as the Fourier series
\begin{equation}
\label{uas}
p_{\rm S}(x;t) = \sum_{k=-\infty}^{+\infty}u_{k}(x)\exp(- {\rm i} k \omega t )\,\,.
\end{equation}
The coefficients $u_{k}(x)$ were found in \cite{varyslope2}. Using the standard matrix notation, they assume the form
\begin{equation}
\label{uk}
u_{k}(x) = m\gamma \frac{|F_{0}|}{D}
\left< k\,|\, \mathbb{L}\,\mathbb{E}(x)\mathbb{R}_{+} \,|\, f \right>\,\,,
\end{equation}
where $\left|f \right>$ is the column vector of the complex amplitudes, $f_{k} = \left< k\,|\,  \mathbb{R}_{-}^{-1} \,|\,  0 \right>$, and the
matrices $\mathbb{E}(x),\, \mathbb{L},\, \mathbb{R}_{\pm}$ possess the matrix elements \cite{varyslope2}
\begin{eqnarray}
\label{Ematrix}
&&\left< m \,|\, \mathbb{E}(x) \,|\, n \right>  =
\frac{\delta_{n\,m}}{2} \left[1+ \frac{1}{\sqrt{1-{\rm i}m \zeta}} \right] 
\exp\! \left[ -\frac{x |F_{0}|}{2 m \gamma D}\left(\sqrt{1-{\rm i}m \zeta} +1\right) \right]\,\,,\\
\label{Lmatrix}
&&\left< m \,|\, \mathbb{L} \,|\, n \right> =
\mathrm{I}_{|m-n|}(-\kappa \sqrt{1-{\rm i} n \zeta}-\kappa)\,\,, \\
\label{Rmatrix}
&&\left< m \,|\, \mathbb{R}_{\pm} \,|\, n \right> =
\mathrm{I}_{|m-n|}(\pm \kappa \sqrt{1-{\rm i} m \zeta}+\kappa)\,\,,
\end{eqnarray}
where $m$ and $n$ are integers, $\kappa = |F_{0}|F_{1}/[2\omega D (m\gamma)^{2}]$, $\zeta= 4\omega D(m\gamma)^{2}/F_{0}^{2}$, and $\mathrm{I}_{k}(\cdot)$ stands for the
modified Bessel function of the first kind, of the order $k$.
Finally, introducing the coefficients
\begin{eqnarray}
\label{ukL}
l_{k}(x) =
\frac{|F_{0}|}{m \gamma D}
\int_{x}^{+\infty}\!\!\!\!\!\! {\rm d}x' \left< k\,|\, \mathbb{L}\, \mathbb{E}(x') \mathbb{R}_{+} \,|\, f \right>\,\,, \\
r_{k}(x) =
\frac{|F_{0}|}{m \gamma D} \int_{0}^{x}{\rm d}x'
\left< k\,|\, \mathbb{L}\, \mathbb{E}(x') \mathbb{R}_{+} \,|\, f \right>\,\,,
\label{ukR}
\end{eqnarray}
the marginal densities (\ref{PleftAS}), (\ref{PrightAS}) can be written in the forms
\begin{eqnarray}
\label{pLconvloution}
p_{{\rm L}}(x;t) =
2\!\!\!\! \sum_{k, n=-\infty}^{+\infty}\!\!\!\! u_{k-n}(x)\, l_{n}(x) \exp(- {\rm i} k \omega t )\,\,, \\
\label{pRconvloution}
p_{{\rm R}}(x;t) =
2\!\!\!\! \sum_{k, n=-\infty}^{+\infty}\!\!\!\! u_{k-n}(x)\, r_{n}(x) \exp(- {\rm i} k \omega t )\,\,.
\end{eqnarray}
These formulae, together with Eq.\ (\ref{uas}), represent an exact asymptotic result and they form the basis of the further discussion.
In the numerical illustrations (Fig.\ 1) we had to reduce the infinite matrices (\ref{Ematrix}), (\ref{Lmatrix}), (\ref{Rmatrix}) into their
(finite) central blocks.

\emph{Mean position} represents the centre of mass of the probability density. In the steady state, we define mean positions of the individual particles as
\begin{equation}
\label{mu}
\mu_{\alpha}(t) = \int_{0}^{+\infty}\!\! {\rm d}x\, x\,  p_{\alpha}(x;t) \,\,, \quad \alpha = {\rm S}, {\rm L}, {\rm R}\,\,.
\end{equation}
Due to the oscillatory driving (\ref{oscildrift}), the mean positions will oscillate with the fundamental frequency $\omega$.
In an average sense, the hard-core interaction produces a repulsive force among the particles. It (in average) shifts the right (left) particle to the right (left) as compared
to the case without the interaction, i.e., $\mu_{\rm L}(t) < \mu_{\rm S}(t) < \mu_{\rm R}(t)$ holds at any instant $t$.

\begin{figure}[t!]
\includegraphics[width=0.7\linewidth]{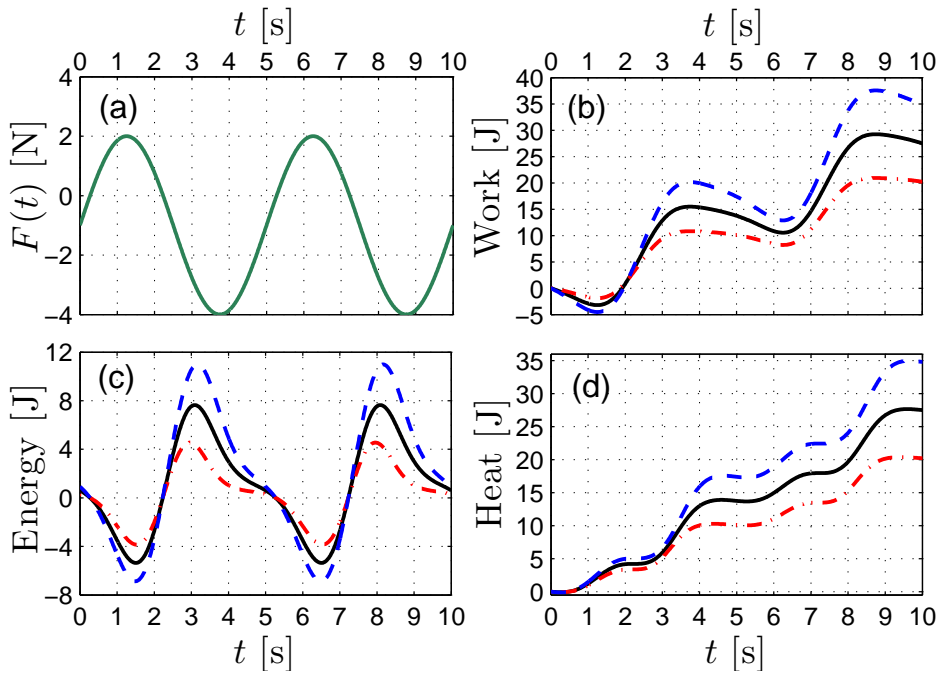}
\caption{The driving force (\ref{oscildrift}) (panel (a)); mean works (\ref{meanworks}) (panel (b)); mean energies (\ref{meanenergies}) (panel (c)); and  mean heats (\ref{meanheats}) (panel (d)) as the functions of time in the steady state. 
In (b), (c) and (d), the solid black line, the blue dashed line, and the red dot-dashed line correspond to the single-diffusing particle, the right particle, and the left particle, respectively. We have used
$F_{0} = - 1.0\, {\rm N}$, $F_{1} = 3.0\, {\rm N}$, $\omega = 0.4\,\pi\, {\rm s^{-1}}$,
$D = 1.0\, {\rm m^{2}\,s^{-1}}$, $m \gamma = 1.0\, {\rm kg\, s^{-1}}$.}
\label{obrazek}
\end{figure}

If we average the external potential $\phi(x,t)$ over all possible positions of the particular particle at a given instant we obtain the \emph{mean energy} of that particle:
\begin{equation}
E_{\alpha}(t) = -[F_{0} + F_{1} \sin(\omega t)]{\mu}_{\alpha}(t)\,\,,
\,\, \alpha = {\rm S}, {\rm L}, {\rm R}\,\,,
\label{meanenergies}
\end{equation}
Generally speaking, the internal energies $E_{\rm S}(t)$, $E_{\rm L}(t)$, $E_{\rm R}(t)$ are periodic functions of time with the fundamental period $2\pi/\omega$. Their oscillations express the combine effect of both the periodically modulated heat flow to the surroundings and the periodic exchange of work done on the particle  by an external agent. From Eq. (\ref{sumdens}), it follows that the total mean energy of two interacting particles is equal to the total mean energy of two non-interacting particles, i.e.,
\begin{equation}
E_{\rm L}(t) + E_{\rm R}(t)= 2 E_{\rm S}(t)\,\,.
\end{equation}
In other words, the hard-core interaction does not contribute to the total energy. Therefore, the repulsive force among the particles arises from the purely entropic effect. 
On the other hand, the mean energies of individual interacting particles are changed significantly as compared to the case without interaction (see Fig. \ref{obrazek}). 

\emph{The mean work}
done on the particle by an external agent during the time interval $[0,t]$ reads \cite{Sekimoto}
\begin{equation}
\label{meanworks}
W_{\alpha}(t) = - F_{1}\omega\!\!\int_{0}^{t}\!{\rm d}t' \cos(\omega t') \mu_{\alpha}(t')\,,
\, \alpha = {\rm S}, {\rm L}, {\rm R}.
\end{equation}
The total work done on the system of two interacting particles again equals to the total work done on two non-interacting particles.
\begin{equation}
\label{sumOFworks}
W_{\rm L}(t) + W_{\rm R}(t) = 2 W_{\rm S}(t)\,\,.
\end{equation}
Nevertheless, as was stressed within the discussion of mean positions, the hard-core interaction in average shifts the right (left) tagged particle to the right (left) as compared to the case without interaction. Hence the absolute value of work done on the right (left) particle is always bigger (smaller) than in the case without interaction.
Since in our setting the diffusion is just a non-equilibrium isothermal process, the total work done on the particles per one period $2 \pi/\omega$ is always positive. This work is entirely dissipated into heat. However, if we allow for a temperature modulation and choose an appropriate temperature schedule,
the system can act as a heat engine which converts a part of the heat accepted form the environment into the useful work.
If this is the case, the right particle will perform a bigger work as compared both to the left particle and to the single-diffusing particle.

\emph{The mean heat} released to the environment and the total \emph{entropy} increase per period are intimately related. From the first law of thermodynamics it follows that
\begin{equation}
\label{meanheats}
Q_{\alpha}(t) = - \left[ E_{\alpha}(t) - W_{\alpha}(t) \right] \,\,, \quad \alpha = {\rm S}, {\rm L}, {\rm R}\,\,.
\end{equation}
is the heat dissipated during the time interval $[0,t]$ by the individual particles.
The entropy generated in the environment due to the dissipative motion of the individual particle during the time interval $[0,t]$ is then
$S_{\alpha}(t) = Q_{\alpha}(t)/T$.
Having periodic changes of the internal energy, the heat dissipated during one period is equal to the work done on the system during one period $Q_{\alpha}(2\pi/\omega) = W_{\alpha}(2\pi/\omega)$. Divided by the temperature of the surroundings  we get the total entropy increase per one period $S_{\alpha}(2\pi/\omega) = Q_{\alpha}(2\pi/\omega)/T$, and
again $S_{\rm L}(2\pi/\omega)+S_{\rm R}(2\pi/\omega)=2S_{\rm S}(2\pi/\omega)$. That is, the entropic repulsion stemming from the hard-core interaction does not influence the
total entropy production.

In summary, the global (aggregative) quantities (e.g.\ the total mean energy (24), the total mean work (26), the total mean entropy production per period) evaluated for the system
of interacting particles equal to those for the system of independent particles. Physically, the conclusion derives from the zero range of the interaction among the particles.
The observation holds for a general external driving and for an arbitrary number of particles, $N$.
Contrary to this, the simple contact interaction strongly influences the one particle dynamical characteristics.

\emph{Acknowledgments:} Support of this work by the Ministry of Education of the Czech Republic (project No. MSM 0021620835) 
and by the project SVV-–2010-–261 305 of the Charles University in Prague is gratefully acknowledged.
\bibliography{paper}

\end{document}